\documentclass[9pt,conference]{IEEEtran}
\usepackage{amssymb,amsthm,amsmath,array}
\usepackage{graphicx}
\usepackage[caption=false,font=footnotesize]{subfig}
\usepackage{xspace}
\usepackage[sort&compress, numbers]{natbib}
\usepackage{stmaryrd}
\usepackage{xcolor}
\usepackage{mathtools}
\usepackage{float}
\usepackage{textcomp}
\usepackage{algorithm}
\usepackage{lipsum}

\begin{document}
\title{Distributed Radar Imaging Based on Accelerated ADMM\\
\thanks{This work is supported by Luxembourg National Research Fund (FNR) through the Industrial Fellowship project "RADII", ref 15364040.}
}
\author{\IEEEauthorblockN{Ahmed Murtada\IEEEauthorrefmark{1}, Bhavani Shankar Mysore Rama Rao\IEEEauthorrefmark{1}, Udo Schroeder\IEEEauthorrefmark{2}}
\IEEEauthorblockA{\IEEEauthorrefmark{1}\textit{Interdisciplinary Centre for Security, Reliability and Trust (SnT),}
\textit{University of Luxembourg}\\
\IEEEauthorrefmark{2}\textit{IEE S.A., Luxembourg}\\
\{ahmed.murtada, bhavani.shankar\}@uni.lu, udo.schroeder@iee.lu} }
\maketitle
\begin{abstract}
The ability of widely distributed radar systems to capture diverse spatial scattering properties substantially improves radar imaging performance. Traditional imaging methods leverage regularized optimization techniques to reconstruct sparse images from local sensors and later combine them to create a global image. Alternatively, we proposed in an earlier work a joint reconstruction technique based on two problem formulations according to the optimization framework of the Alternating Direction Method of Multipliers (ADMM). The joint reconstruction of the global image offers faster convergence, flexible implementation, and a general distributed reconstruction framework. However, despite its benefits, ADMM framework still exhibits a slow convergence rate, making its employment in some contexts impractical. In this paper, we introduce a heuristic method to accelerate the convergence of the previously proposed ADMM formulations based on the gradual elimination of the already converged pixels in accordance with a predetermined criterion. In addition to reducing running time, the accelerated implementation offers reduced computational complexity and lower communication cost between the sensors during iterative updates.
\end{abstract}
\section{Introduction}
\label{sec:intro}
Widely distributed radar systems provide high angular resolution and allow for the exploitation of spatial diversity. However, distributed architecture limits the use of traditional imaging techniques which assume an isotropic scattering model because in a widely distributed system targets exhibit an aspect-dependent scattering behavior. This prevents the use of back-projection (BP) and other traditional imaging algorithms to deliver adequate imaging performance \cite{moses_wide-angle_2004}. Thus, when a wide angle aperture is employed, imaging is typically implemented by splitting the entire aperture into sub-apertures and reconstructing their local images -usually through sparse reconstruction techniques- then fuse them into a global image \cite{moses_wide-angle_2004,ash_wide-angle_2014,sanders_composite_2017}.
In earlier works \cite{hu_widely-distributed_2021,murtada_widely_2022}, we proposed to jointly reconstruct a sparse global image directly in a distributed optimization fashion based on two formulations of ADMM, namely consensus ADMM (CADMM) and sharing ADMM (SADMM). We demonstrated that, in addition to the faster convergence that ADMM naturally offers when compared to other proximal gradient methods applied on measurements obtained by individual sub-apertures \cite{tao2015convergence}, ADMM also offers a flexible distributed processing regime. Nevertheless, when high reconstruction accuracy is considered, ADMM may suffer a noticeable slow convergence \cite{boyd_distributed_2011}. To address this problem, we propose a modified version of the proposed ADMM formulations that further leverage the scene sparsity and gradually learn the approximate support of the scene. The modification offers a reduction of the problem size by heuristically eliminating the already converged pixels and focusing the reconstruction on the sub-images containing the scene support. Therefore, lowering the problem's complexity and significantly decreasing its running time in addition to accelerating the convergence rate. While we introduced the modified version for CADMM formulation in \cite{murtada_accelerated_2022}, in this paper, we introduce and show the performance of the modified version applied to SADMM formulation.\par
%
\section{Signal Model}
We consider a system where $Q$ radar sensors are distributed around a scene of interest. Each sensor is equipped with $M$ receiving antennas and a single transmitting antenna. Sensors transmit linear frequency modulated (LFM) chirps of rate $\alpha$ and a bandwidth $BW$ and each is assumed to receive the reflected echo from the scene due to its own transmission only. Since sensors are widely separated, the system model assumes different reflectivity of the scene with respect to each sensor. By discretizing the scene with a uniform grid of $N$ pixels and stacking the measurement of the receiving antennas at the $q^{\text{th}}$ sensor into a vector ${\mathbf{y}}_{q}$, the received signal can be written in a matrix form as 
\begin{equation}
\label{eq:ModelMat}
\mathbf{y}_{q}=\mathbf{A}_{q}{\mathbf{x}}_{q}+\mathbf{w}_{q}\in \mathbb{C}^{KM\times 1},
\end{equation}
where $\mathbf{A}_{q} \in \mathbb{C}^{KM \times N}$ is the system model-based forward operator as defined in \cite{murtada_accelerated_2022}, $K$ is the number of fast time samples, $\mathbf{x}_{q}$ is the magnitude of the reflectivity of the scene towards the $q^{\text{th}}$ sensor where the phase is assumed to be estimated apriori and incorporated in the matrix $\mathbf{A}_{q}$, $q=1 \ldots Q$, and $\mathbf{w}_{q}$ is additive white Gaussian noise.
%
\begin{figure*}[ht!]
	\centering
	\includegraphics[width=7 in]{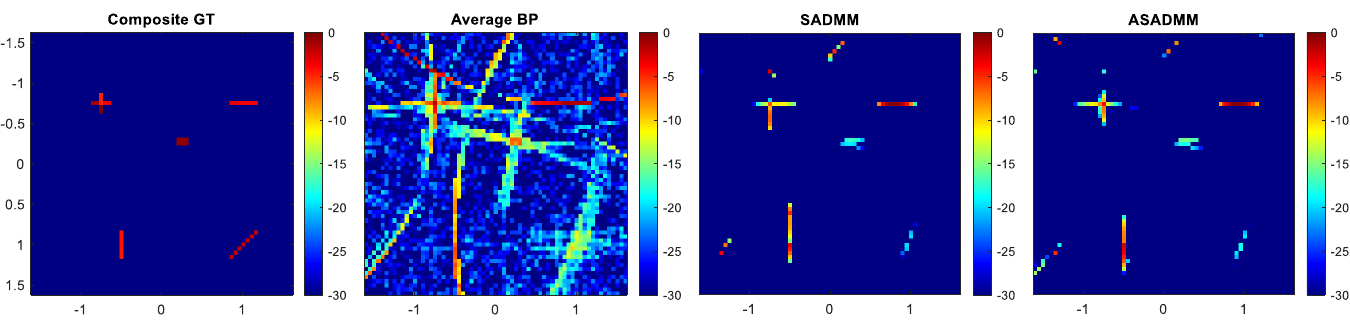}
	\caption{Composite ground truth (GT) image and reconstructed images with a $30$ dB dynamic range using BP, SADMM, and ASADMM }
	\label{fig:ASDMM}
\end{figure*}
\section{Accelerated Sharing ADMM}
As proposed in previous works, we introduce $\mathbf{x_{G}}$ as the variable that represents the magnitude of the global image to be reconstructed. Additionally, we introduce $Q$ matrices $\{\mathbf{P}_{q}\}_{q=1}^{Q}$ of dimensions $N_{q} \times N$ that act as selector operators at the $q^{\text{th}}$ sensor for a sub-image that contains $N_{q}$ pixels out of the full image of the scene. Therefore, the rows are the element selector vectors $\mathbf{e}_{j}^{T} \in \{0,1\}^{N}$ such that $\left[\mathbf{e}_{j}\right]_{i} = \delta_{j-i}$ and $j \in \{1,\ldots,N\}$. \par
Accordingly, the problem formulation according to the accelerated SADMM (ASADMM) formulation is the following
\begin{equation}
\label{eq:SADMM}
\begin{gathered}
\underset{\mathbf{x}_{1},\mathbf{x}_{2},\cdots ,\mathbf{x}_{Q},\mathbf{{x}_{G}}}{\text{arg}\mathop{\min}}\,\, \, \, \,\sum_{q=1}^{Q}{ \frac{\mathrm{\mu}}{2}\left\| \mathbf{y}_{q}-\mathbf{A}_{q} \tilde{\mathbf{P}}_{q} \mathbf{x}_{q} \right\| _{2}^{2}} + \lambda \left\|{\mathbf{{x}_{G}}}\right\|_{1}\\ s.t.\hfill \textstyle \sum_{q=1}^{Q}{\mathbf{P}}_{q}{\mathbf{x}_{q}}-{\mathbf{x}_\mathbf{{G}}}=\mathbf{0}\,\,\,\,\,\,\,\,\,\,\,\,\,\,\hfill
\end{gathered}
\end{equation}
where $\mu$ and $\lambda$ are hyper-parameters to trade-off the data fidelity and ${l}_{1}$ regularization terms and  $\tilde{\mathbf{P}}_{q} = \mathbf{P}_{q}^{T}\mathbf{P}_{q}$.
In addition to the update of the selector matrix that is initialized with identity, the iterative updates of the other variables are obtained by minimizing the augmented Lagrangian function. Consequently, at the $k^{\text{th}}$ iteration we obtain following updates
\begin{equation}
\label{eq:updt_x_SADMM_CF}
\begin{aligned}
    {\hat{\mathbf{x}}}_{q}^{\left(k + 1\right)} &= \resizebox{.85\hsize}{!}{$ \left( \mu \hat{\mathbf{A}}_{q}^{H} \hat{\mathbf{A}}_{q}+\beta \mathbf{I}_{N_{q}} \right)^{-1} \left( \mu \hat{\mathbf{A}}_{q}^{H} \mathbf{y}_{q}+\beta \left({\hat{\mathbf{x}}^{\left(k \right)}_\mathbf{{G}}}-{\sum_{q=1}^{Q}{\mathbf{P}}^{\left(k\right)}_{q}{\hat{\mathbf{x}}_{q}}^{\left(k\right)}} \right) - {\hat{\boldsymbol{\sigma }}}^{\left(k\right)} \right)$}\\
    {\mathbf{{x}}^{\left(k+1 \right)}_\mathbf{{G}}} &=\resizebox{.80\hsize}{!}{$ \mathrm{arg}\mathop{\min} \limits_{{\mathbf{x_{G}}}}\left\{ \lambda \left\| {\mathbf{x_{G}}} \right\|_1 + \frac{\beta}{2}\left\| {\mathbf{x_{G}}} - \bar{\mathbf{x}}^{\left(k + 1\right)} \right\|_{2}^{2} + \hat{\boldsymbol{\sigma}}^{{\left(k \right)}^{T}} {\mathbf{{x}_{G}}}\right\}$}\\
    \hat{\boldsymbol{\sigma }}^{\left(k + 1\right)} &= \hat{\boldsymbol{\sigma}}^{\left(k \right)} + \beta \left( \bar{\mathbf{x}}^{\left(k + 1\right)} - {{\mathbf{x}}}_{\mathbf{{G}}}^{\left(k + 1\right)} \right) \\
    \mathbf{P}_{q}^{\left(k + 1\right)} &= \text{SSC}(\{{\mathbf{x}}_{q}^{\left(i\right)}\}_{i=K_{p}}^{k},\epsilon_{p}) 
\end{aligned}
\end{equation}
where $\beta$ is the augmented penalty parameter, $\hat{\mathbf{x}}_{q} = \mathbf{P}_{q}\mathbf{x}_{q}$, $\hat{\mathbf{A}}_{q} = \mathbf{A}_{q} \mathbf{P}_{q}^{T}$, $\hat{\boldsymbol{\sigma }}$ is the dual variable with the support of the average of all sub-images $\bar{\mathbf{x}}$, and SSC is the subroutine to obtain the updates of the selector matrices detailed in Alg.~\ref{alg:SSC} where $K_p$ and $\epsilon_{p}$ are input parameters for testing convergence that are the number of previous iterations considered and tolerance error, respectively. Note that $\boldsymbol{\eta }_{pri}$ is the primal residual and $\epsilon_{pri}$ is the feasibility tolerance that is obtained based on the user-defined relative and absolute tolerances $\epsilon_{abs}$ and $\epsilon_{rel}$, respectively. The reader may refer to \cite{murtada_widely_2022} for the calculation of these parameters.
\begin{algorithm}[ht!]  \caption{\small $SSC$ subroutine for $ \mathbf{P}_{q}$ update}
\textbf{Input:}  $\left\{{\mathbf{x}}_{q}^{\left(i\right)}\right\}_{i=K_{p}}^{k}$, and $\epsilon_{p}$ \\
\textbf{if} $\left\| \boldsymbol{\eta }^{\left(k\right)}_{pri} \right\|_2 \leq \epsilon_{pri}$
\\
\begin{enumerate}
\item $\mathbf{\zeta} = \frac{1}{K_{p}} \sum_{i=k_i}^{k} |{\mathbf{x}}^{\left(i\right)}-{\mathbf{x}}^{\left(i-1\right)}|$
\item $\{J\} \gets \text{supp}(\mathbf{\zeta} \geq \epsilon_{p})$, $J \in \mathcal{R}^{N_{q}}$
\item $\mathbf{P}_{q} \gets [\mathbf{e}_{J\{1\}}^{T} \mathbf{e}_{J\{2\}}^{T} \cdots \mathbf{e}_{J\{N_q\}}^{T}]^{T}$
\end{enumerate}
\textbf{end if}\\
\textbf{Output:} $\mathbf{P}_{q}$  
\label{alg:SSC}  
\end{algorithm}
%
\section{Numerical Simulations}
We validate the performance of ASADMM on a simulated scene and compare it to SADMM in terms of image reconstruction quality and convergence rate. We run our simulations considering a scene that contains static targets of different shapes that manifest different scattering behavior over the viewing angles. Consequently, targets are assumed to have their radar cross section (RCS) as a function of the viewing angle. Using a fine grid with cells of size $0.05 \ \text{m} \times 0.05 \ \text{m}$, our scene is a region consisting $Nc = 64 \times 64 = 4096$ pixels. To observe the scene, $Q=4$ radar sensors with $M=8$ receiving antennas are distributed around the scene; one per each side. Considering a single sensor transmitting at a time, a linear frequency modulated pulse (LFM) with a sweeping bandwidth $\textrm{BW} = 4\ \text{GHz}$ around the center frequency $f_{c}= 60 \ \text{GHz}$ is used for transmission at all sensors. The received signal is sampled at the Nyquist frequency and corrupted with white Gaussian noise where a signal-to-noise ratio (SNR) of $3$ dB is considered. For image reconstruction, we simulated both SADMM and ASADMM with hyperparameters $\mu = 3$, $\beta = 100$, and $\lambda = 20$ while the absolute and relative tolerance values are set as $\epsilon_{abs} = 10^{-4}$ and $\epsilon_{rel} = 10^{-2}$, respectively. $K_{p}=5$ iterations and $\epsilon_{p}=10^{-5}$ error tolerance are considered for the subroutine SCC in ASADMM implementation.\par
\begin{figure}[h!]
	\centering
	\includegraphics[width=3 in]{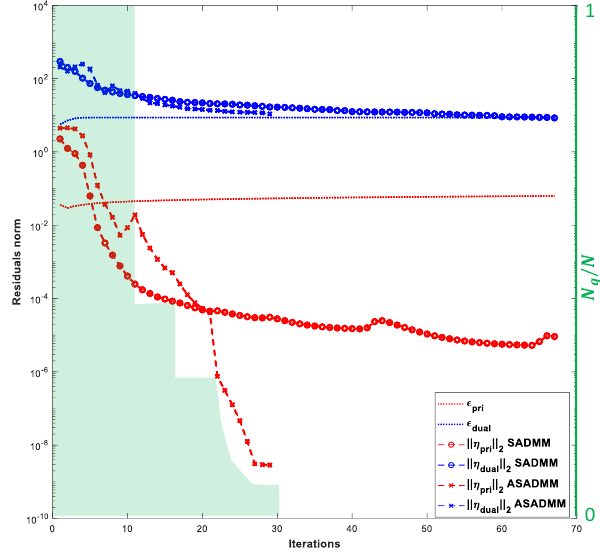}
	\caption{SADMM and ASADMM Convergence Rate}
	\label{fig:Conv}
\end{figure}
 Reconstructed images using SADMM and ASADMM are shown in Fig.~\ref{fig:ASDMM} along with BP reconstruction and the composite GT image. The composite GT is defined as the sum of individual GT images; hence it represents the maximum attainable reconstruction. Fig.~\ref{fig:ASDMM} shows the capability of both algorithms to reconstruct high-quality images of the scene with very similar performance. However, as it can be seen from Fig.~\ref{fig:Conv}, ASADMM requires fewer iterations than needed for SADMM to satisfy the same stopping criteria conditions. Moreover, ASADMM running time is found to be approximately $80 \%$ less than SADMM due to the complexity reduction induced by updating only sub-images of the scene. The maximum sub-image size considered by ASADMM at each iteration is also reported in the background of Fig.~\ref{fig:Conv} where the values are normalized to the full scene size. 
\section{Conclusion}
In this paper, we proposed a modified accelerated version of SADMM algorithm for radar imaging with widely distributed radar sensors. Our proposed algorithm which we called ASADMM provides a similar imaging performance as SADMM reconstructing a high-quality image of the scene. On top of that, ASADMM features a significant reduction in complexity hence much less running time compared to SADMM. The complexity reduction is a result of adaptive scene focusing during the iterative reconstruction of the scene; where ASADMM gradually learns the support of the image portion containing the scattering targets and focuses on that region during the subsequent iterations. Simulation results have verified the performance of ASADMM on synthetic simulated scene. Further investigations regarding performance limitations due to sparsity degree and noise level can be considered in future work.
\clearpage
\bibliographystyle{IEEEtran}
\bibliography{bib.bib}
\end{document}